\documentstyle[prd,preprint,aps,epsf]{revtex}

\begin{document}

\draft

\preprint{$\begin{array}{l}
           \mbox{\bf KEK--TH--422}\\[-3mm]
           \mbox{\bf KEK preprint 94--155}\\[-3mm]
           \mbox{November~1994} \\[-3mm]
           \mbox{\bf H}\\[3cm]
           \end{array}$}

\title{Charged Higgs Effects on Exclusive Semi-tauonic $\bbox{B}$
       Decays}
\author{M.~Tanaka\thanks{Electronic address: minoru@theory.kek.jp}}
\address{Theory Group, KEK, Tsukuba, Ibaraki 305, Japan }
\maketitle
\begin{abstract}
We study effects of charged Higgs boson exchange
in the $B$ semileptonic decays
$\bar B \rightarrow D^{(*)}\tau\bar\nu_\tau$.
Both branching ratio and $\tau$ polarization are examined.
We use the recent experimental data on semileptonic
$B$ decays and the heavy quark effective theory in order to
reduce theoretical uncertainty in the hadronic form
factors. Theoretical uncertainty in the branching ratio
is found to be rather small and that in the $\tau$ polarization
is almost negligible. Their measurements will give
nontrivial constraints on the charged Higgs sector.
\end{abstract}

\vspace{5ex}
\begin{center}
{\em To be published in Zeitschrift f\"ur Physik C.}
\end{center}

\newpage
The evidence for top quarks\cite{CDF} leaves
the Higgs sector as the only
missing part of the standard model (SM). In the minimal SM,
we have one Higgs doublet, which gives a neutral scalar
particle as a physical state. An extension of the Higgs sector is
an interesting possibility for new physics beyond the SM.
One of the most attractive possibilities is the supersymmetric
extension of the SM\cite{HK}. In the minimal supersymmetric SM,
we have to introduce two Higgs doublets in order to cancel
the anomaly and to give the fermions masses. Another important
possibility is CP violation in the Higgs sector\cite{TDL,WB}.
It is known that CP can be broken in the Higgs sector
if we have two or more Higgs doublets.
Since the existence of one or more charged Higgs bosons
is an inevitable consequence of the multi-Higgs-doublet extensions
of the SM, the search for their effects is one of the key points
in the quest for new physics.

The most stringent experimental bound on the charged Higgs boson
mass at present is $m_H\gtrsim 260 {\rm GeV}$, given
by the measurement of the inclusive radiative
$b$ quark decay $b\rightarrow s \gamma$\cite{BSG}.
This bound was found for the two-Higgs-doublet model of
the ``SUSY-type'' Higgs couplings to
fermions, called Model II\cite{HHG}. Since this process
takes place via 1-loop diagrams,
the bound may be changed depending
on details of the model considered. If the model contains
new particles other than the charged Higgs boson which
contribute to the $b\rightarrow s \gamma$ process, the above
lower bound may be modified. This is indeed the case in
the minimal supersymmetric SM\cite{BG}. From this point of view,
it is worthwhile to investigate charged Higgs boson effects in
tree level processes which are less dependent on the other
sectors of the multi-Higgs-doublet models.

In this paper, we study effects of the charged Higgs boson on
the branching ratio and the $\tau$ polarization of the
processes $\bar B \rightarrow D^{(*)}\tau\bar\nu_\tau$.
These processes are expected to be much more sensitive to
the charged Higgs sector than the semileptonic $K$ decay processes
because the Higgs couplings to fermions are proportional
to the fermion mass. The 1\% level branching ratio of these modes
expected in the SM will give of order $10^6$ semi-tauonic
$B$ decay events at the planned $B$ factories.

In $N$-Higgs-doublet models, we have $N-1$ physical
charged Higgs bosons. Their couplings to quarks
and leptons are described by the following interaction
Lagrangian\cite{AST}:
\begin{equation}
{\cal L}_H=(2\sqrt{2}G_F)^{1/2}\sum_{i=1}^{N-1}
            \left[X_i\overline{U}_L V_{KM}M_D D_R+
                  Y_i\overline{U}_R M_U V_{KM}D_L+
                  Z_i\overline{N}_L M_E E_R\right]H_i^+
            +{\rm h.c.}\;.
\label{LAG}
\end{equation}
Here $H_i^{\pm}$ is the i-th lightest physical charged Higgs boson,
\begin{equation}
U_{L(R)}=(u,c,t)_{L(R)}^T\;,\;\;
D_{L(R)}=(d,s,b)_{L(R)}^T\;,\;\;
N_{L}=(\nu_e,\nu_\mu,\nu_\tau)_L^T\;,\;\;
E_R=(e,\mu,\tau)_R^T\;,
\label{FIELDS}
\end{equation}
\begin{equation}
M_U={\rm diag.}(m_u,m_c,m_t)\;,\;\;
M_D={\rm diag.}(m_d,m_s,m_b)\;,\;\;
M_E={\rm diag.}(m_e,m_\mu,m_\tau)\;,
\label{MASSES}
\end{equation}
represent the quark and lepton fields and their masses
respectively, and
$V_{KM}$ is the Kobayashi-Maskawa matrix.
Note that the KM matrix which appears in the charged
current mixing appears in the above Lagrangian.
This is a consequence of the natural
flavor conservation\cite{GW} which we implicitly assumed in
Eq.~(\ref{LAG}) in order to suppress flavor changing neutral
Higgs interactions.

In the case of $N\ge 3$, the coefficients
$X_i,\,Y_i,\,Z_i$ can be complex,
while they are real when $N=2$. In particular they are real
in Model II of
two-Higgs-doublets or in the minimal supersymmetric SM,
and can be written as\cite{HHG}
\begin{equation}
X_1=Z_1=\tan\beta\;,\;\;Y_1=\cot\beta\;,
\label{SC}
\end{equation}
where $\tan\beta=v_u/v_d$ is the ratio of the vacuum expectation
values of the two Higgs fields.

Given the above Lagrangian in Eq.~(\ref{LAG}) and the standard
charged current Lagrangian, we can evaluate effects of the
charged Higgs boson exchange in
$\bar B \rightarrow M\tau\bar\nu_\tau$ processes for
$M=D$ or $D^*$. We adopt a helicity amplitude formalism
since it is convenient for calculating $\tau$ polarizations.
We follow the convention in Refs.~\cite{HMW1} and \cite{HMW2}.

The $W$ boson exchange amplitude is given by
\begin{equation}
{\cal M}_{\lambda_M}^{\lambda_\tau}(q^2,x)_W=
\frac{G_F}{\sqrt{2}}V_{cb} \frac{M_W^2}{M_W^2-q^2}
\sum_{\lambda_W}\eta_{\lambda_W} L_{\lambda_W}^{\lambda_\tau}
H_{\lambda_W}^{\lambda_M}\;,
\label{WEX}
\end{equation}
where $\lambda_M=\pm,0$ denote three possible $D^*$ helicity
states, $\lambda_M=s$ stands for the $D$ mode, and
$\lambda_\tau=\pm$ is the $\tau$ helicity.
The invariant mass squared of the leptonic system is $q^2$, and
$x=p_\tau\cdot p_B/m_B^2$ is the $\tau$ energy divided by
the $B$ meson mass in the $\bar B$ meson rest frame. The virtual
$W$ helicity is denoted by $\lambda_W=\pm,0,s$, and the metric
factor $\eta_{\lambda_W}$ is given by $\eta_{\pm,0}=1$ and
$\eta_s=(q^2-M_W^2)/M_W^2$.
The hadronic and leptonic amplitudes which describe
the processes $\bar B\rightarrow MW^*$ and
$W^*\rightarrow \tau\bar\nu_\tau$ are defined respectively by
\begin{equation}
H_{\lambda_W}^{\lambda_M}(q^2)\equiv
\epsilon_\mu^*(\lambda_W)\langle M(p_M,\lambda_M)|
                         \bar c\gamma^\mu(1-\gamma_5)b
                         |\bar B(p_B)\rangle\;,
\label{HAW}
\end{equation}
and
\begin{equation}
L_{\lambda_W}^{\lambda_\tau}(q^2,x)\equiv
\epsilon_\mu(\lambda_W)\langle\tau(p_\tau,\lambda_\tau)
                      \bar\nu_\tau(p_\nu)|
               \bar \tau\gamma^\mu(1-\gamma_5)\nu_\tau
               |0\rangle\;,
\label{LAW}
\end{equation}
where $\epsilon_\mu(\lambda_W)$ is the polarization vector of
the virtual $W$ boson. Note that the leptonic amplitude
$L_{\lambda_W}^{\lambda_\tau}$ depends on the frame in which
the $\tau$ helicity is defined. $L_{\lambda_W}^{\lambda_\tau}$
with the $\tau$ helicity defined in the virtual $W$ rest frame
is given in Ref.~\cite{HMW1} and the one with the $\tau$ helicity
defined in the initial $\bar B$ rest frame is given in
Ref.~\cite{HMW2}. The hadronic amplitude is also given in
Ref.~\cite{HMW1}.

The amplitude of the charged Higgs exchange can be written as
\begin{equation}
{\cal M}_{\lambda_M}^{\lambda_\tau}(q^2,x)_H=
\frac{G_F}{\sqrt{2}}V_{cb}\sum_i L^{\lambda_\tau}
\left[X_i Z_i^* \frac{m_b m_\tau}{M_{H_i}^2-q^2}H_R^{\lambda_M}+
       Y_i Z_i^* \frac{m_c m_\tau}{M_{H_i}^2-q^2}H_L^{\lambda_M}
\right]\;,
\label{HEX}
\end{equation}
where
\begin{equation}
H_{R,L}^{\lambda_M}(q^2)\equiv
2\langle M(p_M,\lambda_M)|\bar c P_{R,L}b|\bar B(p_B)\rangle\;,
\quad P_R=\frac{1+\gamma_5}{2}\;,\;\;P_L=\frac{1-\gamma_5}{2}\;,
\label{HAH}
\end{equation}
\begin{equation}
L^{\lambda_\tau}(q^2,x)\equiv
\langle\tau(p_\tau,\lambda_\tau)\bar\nu_\tau(p_\nu)|
\bar \tau (1-\gamma_5)\nu_\tau|0\rangle\;.
\label{LAH}
\end{equation}
Using the equations of motion, the hadronic and leptonic amplitudes
of charged Higgs exchange are related to those of $W$ exchange
with scalar $W^*$ polarization ($\lambda_W=s$):
\begin{equation}
H_{R,L}^s=\frac{\sqrt{q^2}}{m_b-m_c} H_s^s\;,\;\;
H_R^\pm=0\;,\quad H_R^0=\frac{\sqrt{q^2}}{m_b+m_c} H_s^0\;,\;\;
H_L^{\pm,0}=-H_R^{\pm,0}\;,
\label{HH}
\end{equation}
\begin{equation}
L^{\lambda_\tau}=\frac{\sqrt{q^2}}{m_\tau}L_s^{\lambda_\tau}\;.
\label{LL}
\end{equation}
As can be seen in Eq.~(\ref{HH}),
the Higgs exchange does not contribute to the decay into
transversely polarized  $D^*$ meson ($\lambda_M=\pm$) because of
angular momentum conservation. Therefore, we study
charged Higgs boson effects in
$\bar B$ decays into $D$ mesons ($\lambda_M=s$)
and those into longitudinally polarized $D^*$ mesons
($\lambda_M=0$) in the $\bar B$ rest frame.

Using the helicity amplitudes of Eqs.~(\ref{WEX}) and (\ref{HEX}),
it is straightforward to calculate the differential decay rate%
\footnote{The decay distributions of the charge conjugate processes
          ($B\rightarrow\bar M\tau^+\nu_\tau$ with
           $M=\bar D,\;\bar D^*$) are obtained by
          taking complex conjugate of all the couplings.}:
\begin{equation}
d\Gamma_{\lambda_M}=
 \frac{1}{2 m_B}\sum_{\lambda_\tau}
 |{\cal M}_{\lambda_M}^{\lambda_\tau}|^2
 d\Phi_3\;,
\label{DR}
\end{equation}
where ${\cal M}_{\lambda_M}^{\lambda_\tau}=
       {\cal M}_{\lambda_M}^{\lambda_\tau}(q^2,x)_W+
       {\cal M}_{\lambda_M}^{\lambda_\tau}(q^2,x)_H$,
and $d\Phi_3=dq^2\,dx/64\pi^3$ is the three-body phase space.
Also, $\tau$ polarizations can be calculated conveniently
with the helicity amplitudes. Let us consider the decay rate
with a definite $\tau$ spin direction. It can be written as
\begin{equation}
d\Gamma_{\lambda_M}(\bbox{s})=
 \frac{1}{2}\left[d\Gamma_{\lambda_M}+
 (d\Gamma_{\lambda_M}^L\bbox{e}_L+
  d\Gamma_{\lambda_M}^\bot \bbox{e}_\bot+
  d\Gamma_{\lambda_M}^T \bbox{e}_T)\cdot\bbox{s}
            \right]\;,
\label{DRS}
\end{equation}
where $\bbox{s}$ is the unit vector which points toward the $\tau$
spin direction in the $\tau$ rest frame, and the basis vectors
are defined as
$\bbox{e}_L\equiv\bbox{p}_\tau/|\bbox{p}_\tau|$,
$\bbox{e}_T\equiv\bbox{p}_M\times\bbox{p}_\tau/
                 |\bbox{p}_M\times\bbox{p}_\tau|$,
and $\bbox{e}_\bot\equiv\bbox{e}_T\times
                        \bbox{e}_L$,
with the convention that the angle from $\bbox{p}_M$ to
$\bbox{p}_\tau$ lies between $0$ and $\pi$. The situation
is depicted in Fig.~\ref{KD}.
These definitions can be used both in the virtual $W$ (or Higgs)
rest frame and in the $\bar B$ rest frame.
The components of the spin-dependent part of the decay rate
in Eq.~(\ref{DRS}) are given in terms of the helicity
amplitudes as
\begin{equation}
d\Gamma_{\lambda_M}^L=
  \frac{1}{2 m_B}\left(|{\cal M}_{\lambda_M}^+|^2-
                       |{\cal M}_{\lambda_M}^-|^2\right)
   d\Phi_3\;,
\label{DRL}
\end{equation}
\begin{equation}
d\Gamma_{\lambda_M}^\bot=
  \frac{1}{m_B}{\rm Re}\left({\cal M}_{\lambda_M}^{+*}
                       {\cal M}_{\lambda_M}^-\right)
   d\Phi_3\;,
\label{DRP}
\end{equation}
\begin{equation}
d\Gamma_{\lambda_M}^T=
  \frac{1}{m_B}{\rm Im}\left({\cal M}_{\lambda_M}^{+*}
                       {\cal M}_{\lambda_M}^-\right)
   d\Phi_3\;.
\label{DRT}
\end{equation}
{}From Eq.~(\ref{DRS}), the possible three
$\tau$ polarizations are defined by
\begin{equation}
P_L\equiv \frac{d\Gamma(\bbox{e}_L)-
                d\Gamma(-\bbox{e}_L)}
               {d\Gamma(\bbox{e}_L)+
                d\Gamma(-\bbox{e}_L)}
   =\frac{ d\Gamma^L}{d\Gamma}\;,
\label{PL}
\end{equation}
\begin{equation}
P_\bot\equiv \frac{d\Gamma(\bbox{e}_\bot)-
                   d\Gamma(-\bbox{e}_\bot)}
                  {d\Gamma(\bbox{e}_\bot)+
                   d\Gamma(-\bbox{e}_\bot)}
   =\frac{ d\Gamma^\bot}{d\Gamma}\;,
\label{PP}
\end{equation}
\begin{equation}
P_T\equiv \frac{d\Gamma(\bbox{e}_T)-
                d\Gamma(-\bbox{e}_T)}
               {d\Gamma(\bbox{e}_T)+
                d\Gamma(-\bbox{e}_T)}
   =\frac{ d\Gamma^T}{d\Gamma}\;,
\label{PT}
\end{equation}
where we omit the common helicity index $\lambda_M$.
Note that $P_L^2+P_\bot^2+P_T^2=1$ at an arbitrary kinematical
configuration because of the definite neutrino helicity.
The longitudinal polarization ($P_L$) and the perpendicular
polarization ($P_\bot$) depend on the frame in which the $\tau$
helicity is defined. On the other hand, the transverse
polarization ($P_T$) is frame-independent.
It is well-known that the transverse polarization is a
T-violating quantity as long as the final state interaction
can be ignored. The T- or CP-violating nature of the transverse
polarization can be seen in Eq.~(\ref{DRT}) since
all the tree-level amplitudes of Eqs.~(\ref{WEX}) and
(\ref{HEX}) are chosen to be real in the CP-conserving
limit in our convention.

In order to find numerical predictions, the hadronic transition
form factors are needed. In Ref.~\cite{HMW1}, the following set
of hadronic form factors are employed and the hadronic amplitude
$H_{\lambda_W}^{\lambda_M}$ in Eq.~(\ref{HAW}) is given
in terms of them:
\begin{equation}
\langle D(p_M)|\bar c\gamma^\mu b|\bar B(p_B)\rangle=
 f_+(q^2)(p_B+p_M)^\mu+f_-(q^2)(p_B-p_M)^\mu\;,
\label{FD}
\end{equation}
\begin{equation}
\langle D^*(p_M,\lambda_M)|\bar c\gamma^\mu b|\bar B(p_B)\rangle=
 if_1(q^2)\epsilon^{\mu\nu\rho\sigma}
 \epsilon^*_{M\nu}(p_B+p_M)_\rho(p_B-p_M)_\sigma\;,
\label{FDSV}
\end{equation}
\begin{equation}
\langle D^*(p_M,\lambda_M)|\bar c\gamma^\mu\gamma_5 b
  |\bar B(p_B)\rangle=
 f_2(q^2)\epsilon^{*\mu}_M+\epsilon^*_M\cdot p_B
 \left\{f_3(q^2)(p_B+p_M)^\mu+f_4(q^2)(p_B-p_M)^\mu\right\}\;,
\label{FDSA}
\end{equation}
where $\epsilon_M=\epsilon_M(p_M,\lambda_M)$ is
the polarization vector of the $D^*$ meson.
In the following, however, we adopt a different set of form factors
which is more convenient to incorporate the results of the heavy
quark effective theory. We use the following form factors\cite{N1}:
\begin{equation}
\langle D(v')|\bar c\gamma^\mu b|\bar B(v)\rangle=
 \sqrt{m_Bm_M}\left[\xi_+(y)(v+v')^\mu+
                    \xi_-(y)(v-v')^\mu\right]\;,
\label{HFD}
\end{equation}
\begin{equation}
\langle D^*(v',\lambda_M)|\bar c\gamma^\mu b|\bar B(v)\rangle=
 i\sqrt{m_Bm_M}\,\xi_V(y)\epsilon^{\mu\nu\rho\sigma}
 \epsilon^*_{M\nu}v'_\rho v_\sigma\;,
\label{HFDSV}
\end{equation}
\begin{equation}
\langle D^*(v',\lambda_M)|\bar c\gamma^\mu\gamma_5 b
  |\bar B(v)\rangle=\sqrt{m_Bm_M}\left[
 \xi_{A_1}(y)(1+y)\epsilon^{*\mu}_M-
 \xi_{A_2}(y)\epsilon^*_M\cdot v\; v^\mu-
 \xi_{A_3}(y)\epsilon^*_M\cdot v\; v'^\mu\right]\;,
\label{HFDSA}
\end{equation}
where $v=p_B/m_B$ and $v'=p_M/m_M$ are the four-velocities
of the $\bar B$ and $M(=D,D^*)$ mesons respectively, and
$y\equiv v\cdot v'=(m_B^2+m_M^2-q^2)/(2 m_B m_M)$.
The form factors in Eqs.~(\ref{FD})$\sim$(\ref{FDSA}) can be
written in terms of the form factors in
Eqs.~(\ref{HFD})$\sim$(\ref{HFDSA}):
\begin{equation}
f_\pm=\pm\frac{1}{2\sqrt{r}}[(1\pm r)\xi_+-(1\mp r)\xi_-]\;,
\label{FR1}
\end{equation}
\begin{equation}
f_1=\frac{1}{2m_B\sqrt{r}}\,\xi_V\;,\;\;
f_2=m_B\left(\sqrt{r}+\frac{p_B\cdot p_M}{m_B^2\sqrt{r}}\right)
    \xi_{A_1}\;,\;\;
f_{3,4}=-\frac{1}{2m_B}\left(\sqrt{r}\,\xi_{A_2}\pm
                         \frac{1}{\sqrt{r}}\,\xi_{A_3}\right)\;,
\label{FR2}
\end{equation}
where $r=m_M/m_B$ and we omitted the arguments of
the $f\/$'s and $\xi\/$'s.

In the heavy quark limit and in the leading logarithmic
approximation (LLA), we have\cite{QCD,IW}
\begin{equation}
\xi_+=\xi_V=\xi_{A_1}=\xi_{A_3}\equiv C\,\xi\;,\;\;
\xi_-=\xi_{A_2}=0\;,
\label{HQL}
\end{equation}
where $\xi(1)=1$ and $C$ denotes the QCD correction factor
in the LLA%
\footnote{The factor $C$ drops out in our results.}.
We assume the following form of the universal form factor%
\footnote{The following results in this paper are not affected
          significantly by adopting alternative
          forms such as those used in Ref.~\protect\cite{ARGUS}},

\begin{equation}
\xi(y)=\left(\frac{2}{1+y}\right)^{2\rho^2}\;,
\label{POLE}
\end{equation}
and determine the slope parameter $\rho$ from the experimental
data of semileptonic $B$ decays\cite{CLEO}. As a result, we obtain
\begin{equation}
\rho=1.08\pm 0.11\;,
\label{RHQL}
\end{equation}
with $\chi^2_{min}/\mbox{\em d.o.f.}=0.48$.
Eq.~(\ref{RHQL}) gives the uncertainty of the predictions
in the approximation of Eqs.~(\ref{HQL}) and (\ref{POLE}).

In our numerical analysis, we consider only the effects of
the lightest charged Higgs boson exchange.
Other charged Higgs bosons (if they exist)
are assumed to be too heavy to give significant contributions.
Moreover, we concentrate on the case of the Model II Higgs
couplings as shown in Eq.~(\ref{SC}). So, the strength of the
charged Higgs couplings to fermions are determined by
$\tan\beta$ only.

We use $m_b=4.8$GeV, $m_c=1.4$GeV, and $m_\tau=1.78$GeV, and
we do not consider the uncertainty in the quark masses,
because its effect appears mostly through the combination
$m_bm_\tau\tan^2\beta/M_H^2$ and a variation in $m_b$ can be
absorbed by changing $\tan\beta$ or the charged Higgs
boson mass $M_H$.

Our predictions for the branching ratio are shown in
Figs.~\ref{BD} and \ref{BDL}.
In Figs.~\ref{BD}(a) and \ref{BDL}(a),
we show the decay rate of the process
$\bar B \rightarrow D^{(*)}\tau\bar\nu_\tau$ in the presence
of the charged Higgs boson exchange, normalized to the rate of
$\bar B \rightarrow D^{(*)}\mu\bar\nu_\mu$ in the SM,
against the charged Higgs mass for several values of $\tan\beta$.
The shaded regions correspond to the predictions in the
approximation of Eqs.~(\ref{HQL}) and (\ref{POLE})
within the uncertainty of Eq.~(\ref{RHQL}).
In Figs.~\ref{BD}(b) and \ref{BDL}(b),
we also show the decay rate normalized to $\tilde\Gamma$,
the decay rate of
$\bar B \rightarrow D^{(*)}\mu\bar\nu_\mu$ in the SM, but
integrated in the same $q^2$ region as the $\tau$ mode,
{\em i.e.\/} $m_\tau^2\leq q^2\leq (m_B-m_M)^2$.
As seen in Figs.~\ref{BD} and \ref{BDL},
this restriction in the $q^2$ range decreases
the uncertainty in the hadronic form factors
and improve the sensitivity to
the charged Higgs sector. The theoretical sensitivity to
the charged Higgs sector can be represented by the minimum value
of $R\equiv M_W\tan\beta/M_H$ which can be detected
in an ideal experiment. From Fig.~\ref{BD}(b), we expect
the theoretical reach of $R\sim 6$ with the uncertainty
of Eq.~(\ref{RHQL}).
As can be seen in Fig.~\ref{BDL}, the longitudinal
$D^*$ ($D^*_L$) mode is less
sensitive to the charged Higgs boson exchange because of
the angular momentum barrier. Actually, the hadronic amplitude
$H_s^0(q^2)$ vanishes as $q^2\rightarrow (m_B-m_M)^2$.

In order to get an idea on the size of
the $1/m_{b,c}$ correction and the non-LLA QCD correction which
violate the relation among the form factors in Eq.~(\ref{HQL}),
we employ the estimation of these corrections by
Neubert\cite{N2}. In Ref.~\cite{N2} both
the $1/m_{b,c}$ correction as estimated by the QCD sum rule
and the perturbative QCD correction beyond the LLA
are given. The form factors in
Eqs.~(\ref{HFD})$\sim$(\ref{HFDSA}) are then written as
\begin{equation}
\xi_i(y)=\left[\alpha_i+\beta_i(y)+
                       \gamma_i(y)\right]\xi(y)\;,
\qquad i=+,-,V,A_1,A_2,A_3\;,
\label{FC}
\end{equation}
where $\alpha_+=\alpha_V=\alpha_{A_1}=\alpha_{A_3}=1$,
$\alpha_-=\alpha_{A_2}=0$, $\beta_i(y)$ represents
the perturbative QCD correction, and $\gamma_i(y)$ is the
$1/m_{b,c}$ correction. The functions $\beta_i$'s and
$\gamma_i$'s are given in Ref.~\cite{N2}. Assuming
the form of $\xi(y)$ in Eq.~(\ref{POLE}) again,
we obtain the following range of the slope parameter from
the experimental data\cite{CLEO}:
\begin{equation}
\rho=1.23\pm 0.09\;,
\label{RC}
\end{equation}
with $\chi^2_{min}/\mbox{\em d.o.f.}=0.55$.

The results on the decay rate by using Eq.~(\ref{FC})
and the central value of Eq.~(\ref{RC}), {\em i.e.\/}
$\rho=1.23$, are shown in Figs.~\ref{BD} and \ref{BDL}
by dashed lines.
The magnitude of the uncertainty
from the range of the parameter $\rho$ in Eq.~(\ref{RC})
is roughly as the same as
in the leading order approximation of
Eqs.~(\ref{HQL}) $\sim$ (\ref{RHQL}). Fig.~\ref{BD} shows that
the non-leading corrections are not the major uncertainty
in the $\bar B$ to $D$ mode, and the theoretical reach of
$R\sim 6$ for the $D$ mode remains valid.
On the other hand, as can be seen from Fig.~\ref{BDL},
the non-leading corrections can be as large as the uncertainty
in the slope parameter of Eq.~(\ref{RC}) for the $D^*_L$ mode.
In future, the non-leading corrections may become dominant
uncertainties in both the $D$ and $D^*_L$ modes if
the ranges of Eqs.~(\ref{RHQL}) and (\ref{RC}) are
reduced enough by detailed studies of the semileptonic
$B$ decays at $B$ factories.

As for the $\tau$ polarizations, we can calculate any of
their distributions by using Eqs.~(\ref{DR}) and
(\ref{DRL})$\sim$(\ref{PT})
given the helicity amplitudes. For the couplings of
Eq.~(\ref{SC}), we obtain $P_T=0$, and $P_L^2+P_\bot^2=1$ at any
phase space point $(q^2,x)$. However, for simplicity,
we concentrate on the longitudinal polarization $P_L$
integrated over the whole phase space separately in the numerator
and the denominator of Eq.~(\ref{PL}). Note that
$P_L^2+P_\bot^2\neq 1$ after this integration.
In the following, we consider two
of the possible frames in which the $\tau$ helicity is defined,
the virtual $W(H)$ rest frame and the $\bar B$ rest frame.
The longitudinal $\tau$ polarization in the former frame
is denoted by $P_L(W^*)$, and in the latter frame
by $P_L(B)$.

Our numerical results on $P_L(W^*)$ and $P_L(B)$ are
given in Figs.~\ref{PD} and \ref{PDL}.
Fig.~\ref{PD}(a) and (b) show the $P_L(W^*)$ and
the $P_L(B)$ respectively in the $D$ mode.
Fig.~\ref{PDL}(a) and (b) show the same quantities
in the $D^*_L$ mode.
The values in the leading order approximation
of Eqs.~(\ref{HQL}) and (\ref{POLE}) with
the uncertainty due to the range of $\rho$ in
Eq.~(\ref{RHQL}) are again shown by the shaded region.
The uncertainty in the prediction is found to be much smaller
than that in the branching ratio; it is
almost negligible except for $P_L(B)$ in the $D$ mode.
The predictions with the non-leading corrections of Eq.~(\ref{FC})
for the central value of Eq.~(\ref{RC}) are also shown
by dashed lines. The non-leading corrections
can be regarded as the major uncertainty in the calculations
of these polarizations except for $P_L(B)$ in the $D$ mode.

{}From Fig.~\ref{PD}(a),
we expect the best possible theoretical reach of
$R\sim 4.5$ among the calculations in this paper,
regarding the difference between the center
line of the shaded region which shrinks to almost a line
and the dashed line as possible
theoretical uncertainties. However, this estimation of
the theoretical reach is rather ambiguous because it heavily
relies on the estimation of the non-leading corrections
in Ref.~\cite{N2}. On the other hand, as expected,
the $D^*_L$ mode is less sensitive to the charged Higgs boson
exchange in these polarizations too, see Fig.~\ref{PDL}.

The above results on the sensitivity of the branching ratio
and the $\tau$ polarizations to the charged Higgs boson effects
in the exclusive semi-tauonic $B$
decays should be compared with those in the inclusive
semi-tauonic $B$ decay.
The inclusive decay has been studied in Ref.~\cite{GL}, which
finds the sensitivity of $R\sim 32$ for the branching ratio,
and $R\sim 20$ for the $\tau$ polarization%
\footnote{In Ref.~\protect\cite{GL}, the longitudinal $\tau$
polarization in the $\bar B$ rest frame is discussed.},
despite the fact that the uncertainties in the inclusive study of
Ref.~\cite{GL} seem to be slightly smaller than those
in the exclusive study of the present paper.
These less sensitive results
are understood as the above-explained insensitivity
of the $D^*$ mode which gives a large portion of
the inclusive decay and cannot be separated
in the inclusive study. In other words, the $D^*$ modes
dilute the effects of the charged Higgs boson exchange
in the inclusive decay.
On the other hand, in the exclusive study,
we can select out the sensitive
$D$ mode. Therefore, measurements of the branching ratio and
the $\tau$ polarizations for
$\bar B\rightarrow D\tau\bar\nu_\tau$
mode may give the best bound on the charged Higgs boson exchange
at the tree level.

Finally, we comment on reconstruction of the $\tau$ momentum.
Reconstruction of the $\tau$ momentum is desirable in
several measurements discussed above, in particular it
is necessary to measure $x$ and $q^2$ distributions.
Even if we know the momenta $p_B$, $p_M$, and $p_h$
in the decay process $\bar B\rightarrow M\tau\bar\nu_\tau$
followed by $\tau\rightarrow {\rm hadrons}(h)+\nu_\tau$,
the $\tau$ momentum cannot be reconstructed because of
two missing neutrinos. In this case, the $\tau$ momentum
is parametrized by its azimuthal angle in the virtual
$W(H)$ rest frame in which $\bbox{p}_h$ points toward
the positive $z$ direction. Improvements in vertex detector
technology can improve the situation.
Measurements of several quantities in semi-tauonic $B$
decays should be improved significantly by systematically
taking into account the vertex information. In principle,
by using the knowledge about the tracks originating from
$M$ and $\tau$, we can measure the impact parameter between the
flight lines of $M$ and $h$. If the position of
the $\bar B$ decay vertex or the $\tau$ decay vertex
is known in addition to this impact parameter, we can
obtain the $\tau$ momentum by determining the azimuthal
angle mentioned above. However, there remains a
two-fold ambiguity in general. To disentangle this ambiguity,
both the $\bar B$ decay vertex and the $\tau$ decay vertex
should be known. In principle, the $\bar B$ decay vertex
can be measured when the beam axis is well-known
or when $\bar B$ decays into $D^*$.
The $\tau$ decay vertex can be measured
in the three-prong decays.

The author would like to thank K.~Hagiwara for
his careful reading of the manuscript and valuable
discussions. He also thanks Y.~Kuno for useful
discussions and B.~Bullock for his reading of the
manuscript.

\begin{figure}
\caption{Kinematics and definition of the basis vectors of
         $\tau$ polarization.}
\label{KD}
\end{figure}

\begin{figure}
\caption{The branching ratios for the $D$ mode:
         The shaded regions show the predictions within
         the uncertainty in $\rho$ (Eq.~(\protect\ref{RHQL}))
         in the approximation of Eqs.~(\protect\ref{HQL})
         and (\protect\ref{POLE}),
         and the dashed lines are the predictions with
         the non-leading corrections of Eq.~(\protect\ref{FC})
         for the central value of Eq.~(\protect\ref{RC}).
         (a) The decay rate normalized to that of
         $\bar B\rightarrow D\mu\bar\nu_\mu$ in the SM:
         (b) The same as (a) except that the denominator is
         integrated over the region
         $m_\tau^2\leq q^2\leq (m_B-m_M)^2$.}

\label{BD}
\end{figure}

\begin{figure}
\caption{The branching ratios for the $D^*_L$ mode:
         The same as Fig.~\protect\ref{BD}.}
\label{BDL}
\end{figure}

\begin{figure}
\caption{The $\tau$ polarizations for the $D$ mode:
         The uncertainties are shown in the same way
         as Figs.~\protect\ref{BD} and \protect\ref{BDL}.
         The longitudinal $\tau$ polarizations,
         (a) in the virtual $W$($H$) rest frame, and
         (b) in the $\bar B$ rest frame are shown.}
\label{PD}
\end{figure}

\begin{figure}
\caption{The $\tau$ polarizations for the $D^*_L$ mode:
         The same as Fig.~\protect\ref{PD}.}
\label{PDL}
\end{figure}

\newpage
\hspace*{1cm}
\epsfbox{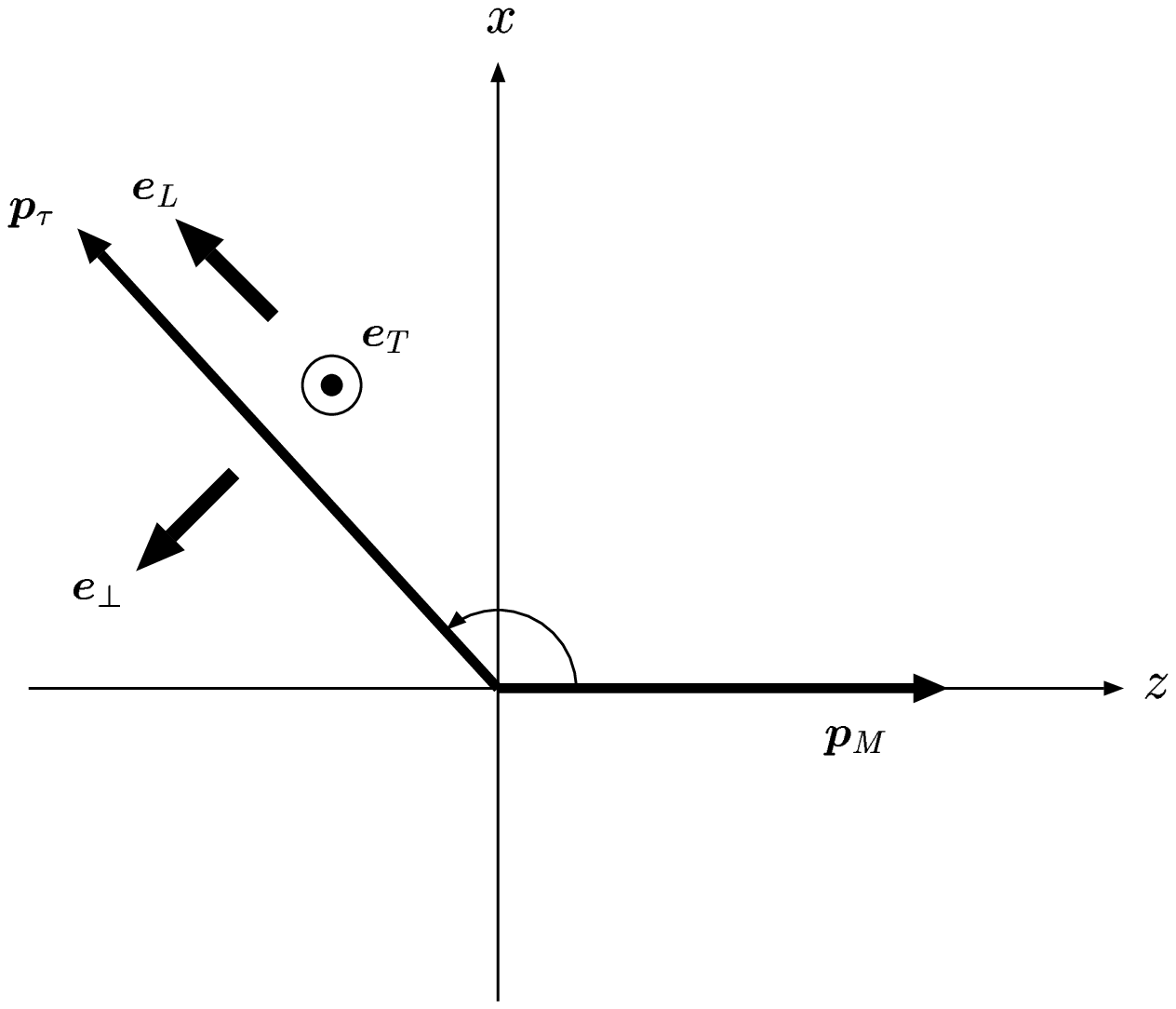}
\begin{center}
{\Large Fig.~\ref{KD}}
\end{center}

\newpage
\epsfbox{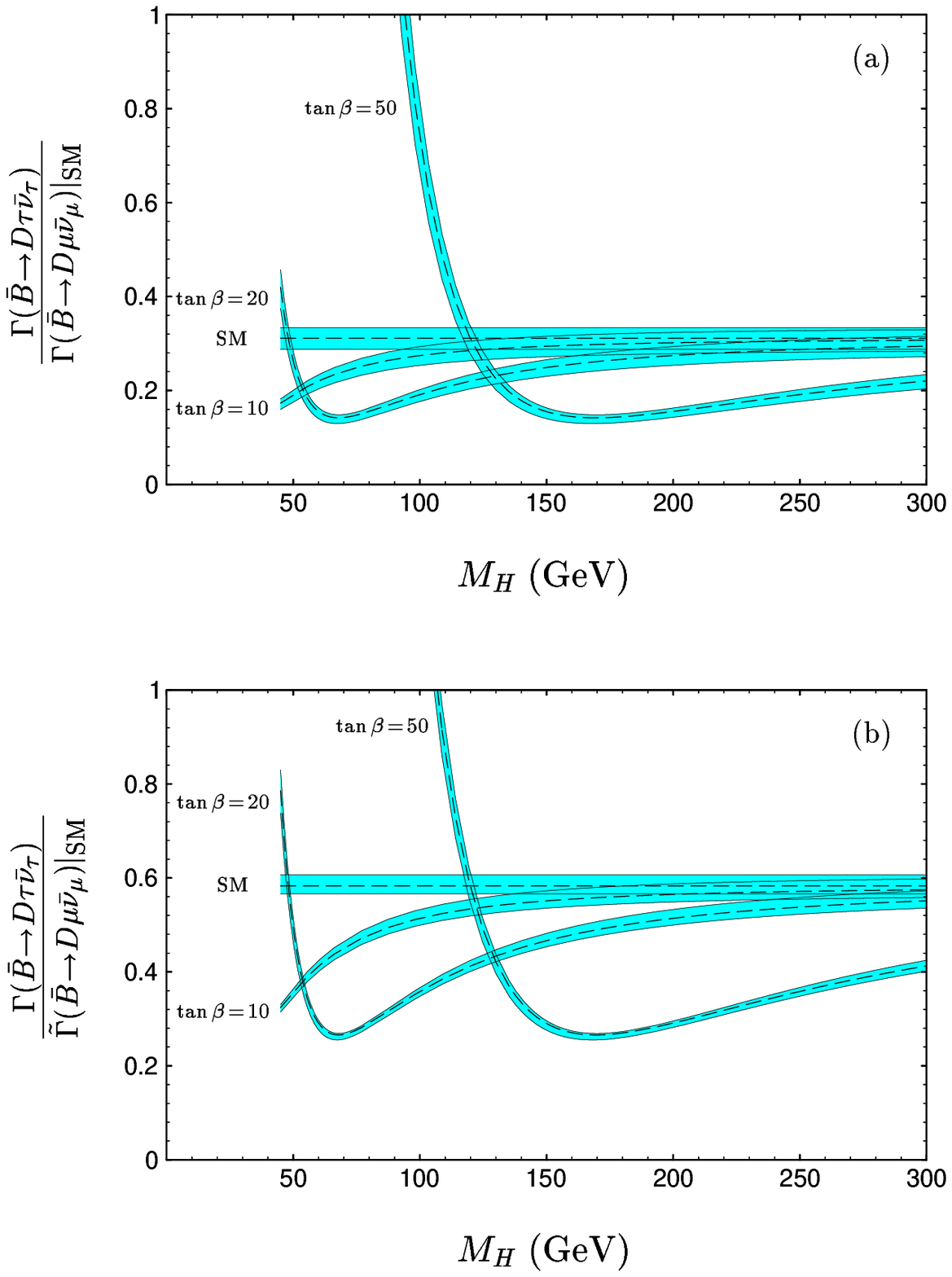}
\begin{center}
{\Large Fig.~\ref{BD}}
\end{center}

\newpage
\epsfbox{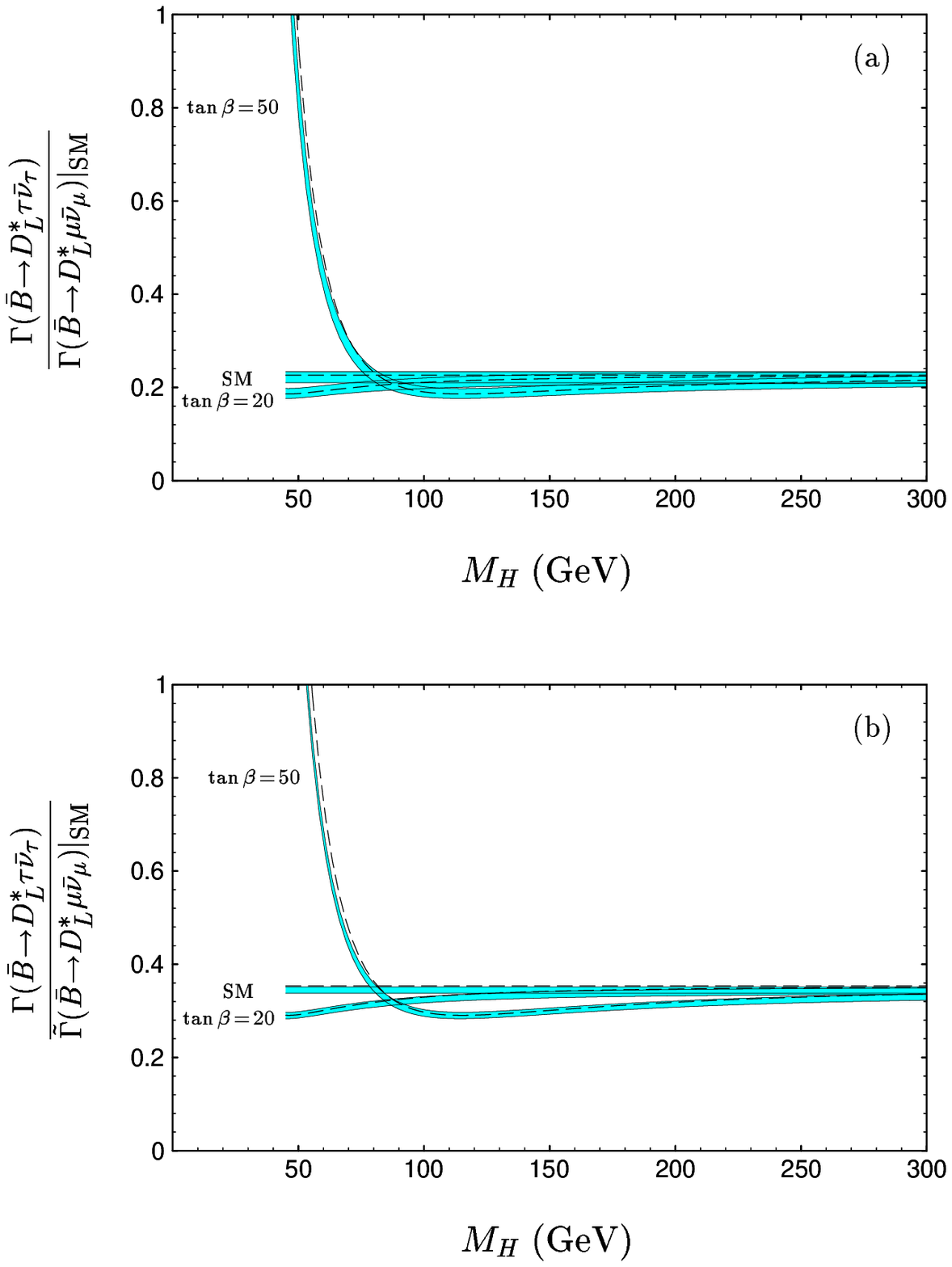}
\begin{center}
{\Large Fig.~\ref{BDL}}
\end{center}

\newpage
\epsfbox{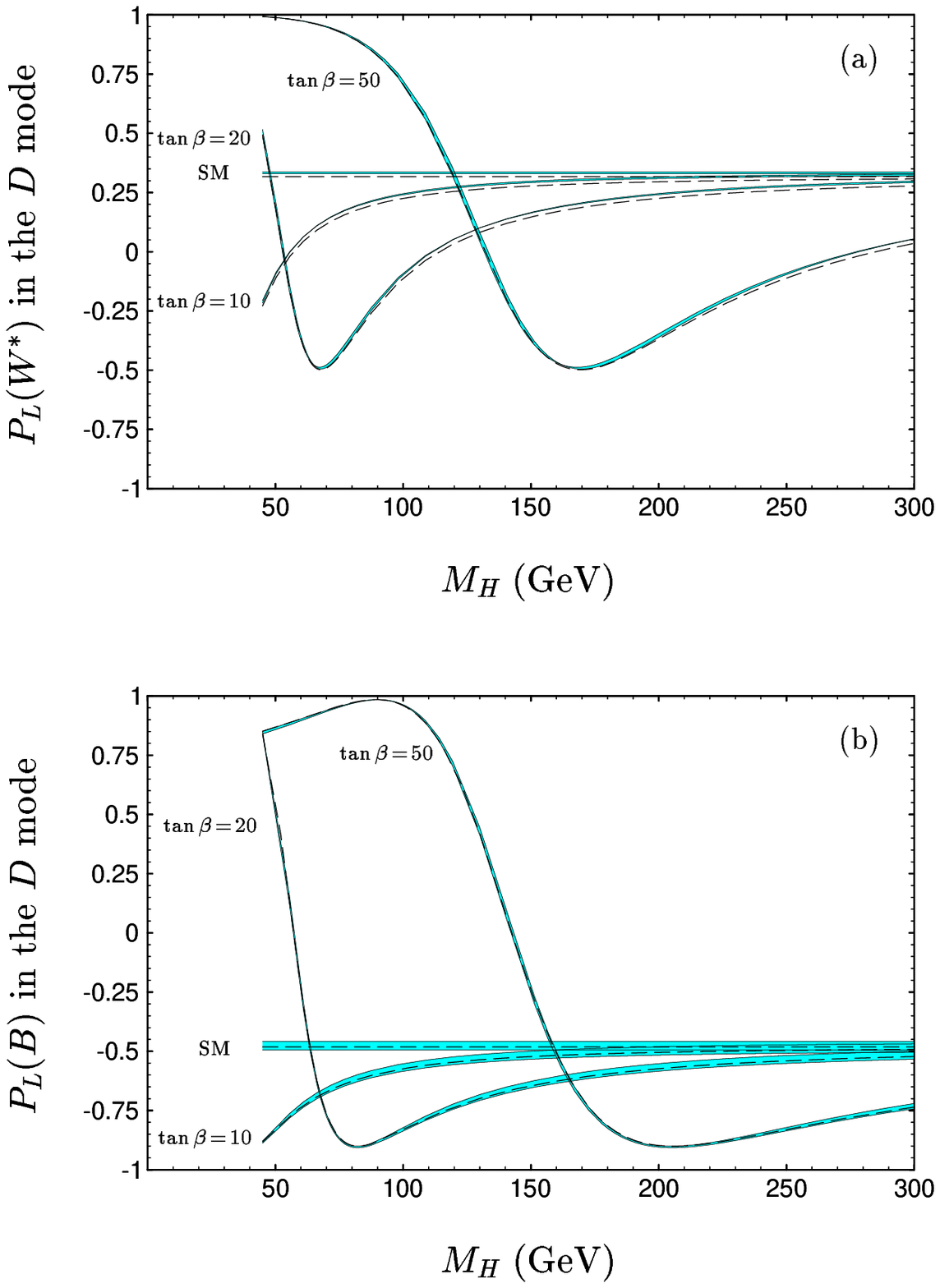}
\begin{center}
{\Large Fig.~\ref{PD}}
\end{center}

\newpage
\epsfbox{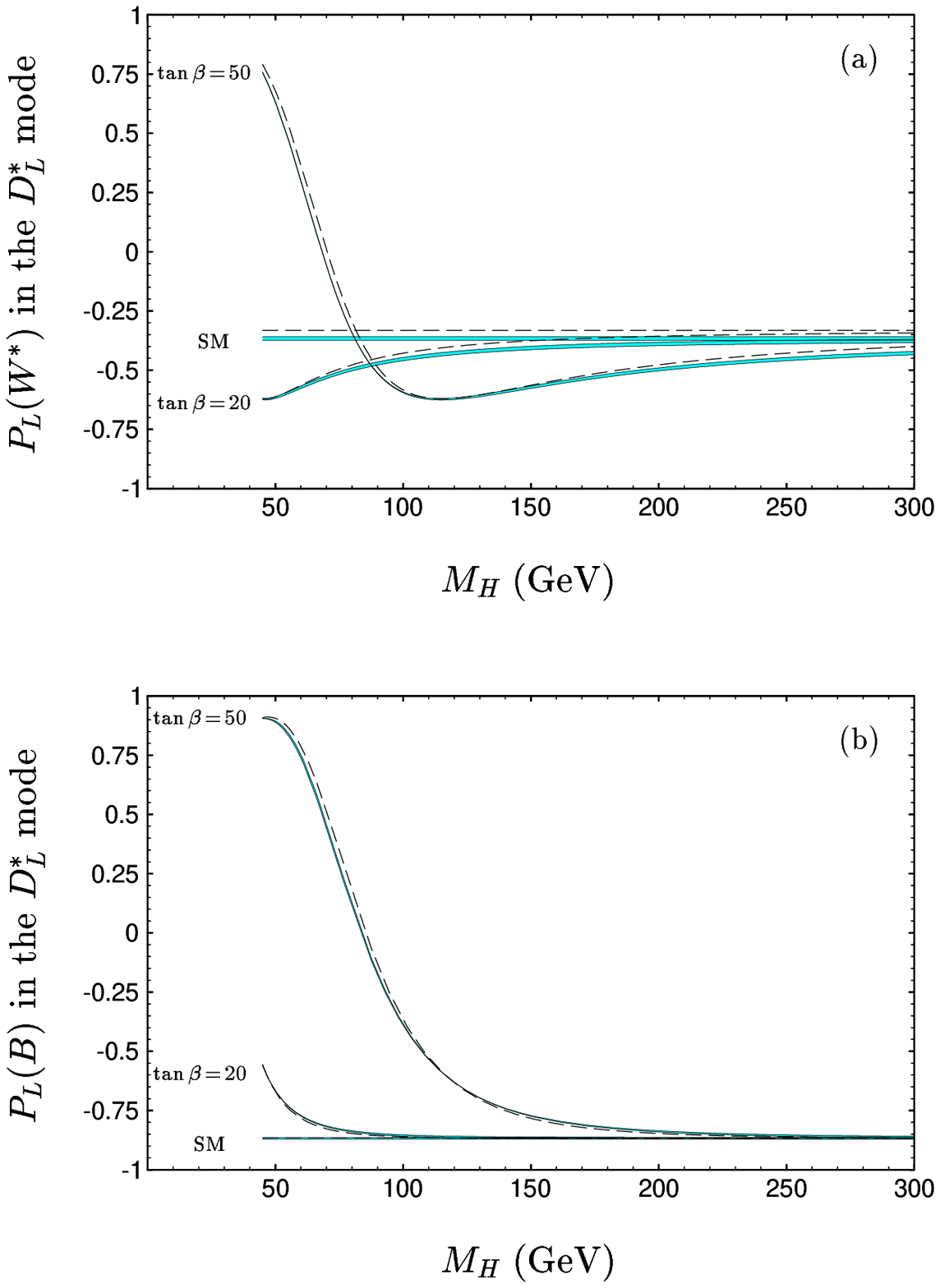}
\begin{center}
{\Large Fig.~\ref{PDL}}
\end{center}

\end{document}